# Confined Hyperbolic Metasurface Modes for Structured Illumination Microscopy

**John Haug [1,*], Milan Palei [1], Joshua D. Shrout [2,3], Evgenii Narimanov [4], Paul W. Bohn [5,6], and Anthony J. Hoffman [1]**

*1 Department of Electrical Engineering, University of Notre Dame, Notre Dame, IN 46556, USA*
*2 Department of Civil and Environmental Engineering and Earth Sciences, University of Notre Dame, Notre Dame, IN 46556, USA*
*3 Department of Biological Sciences, University of Notre Dame, Notre Dame, IN 46556, USA*
*4 Brick Nanotechnology Center, School of Computer and Electrical Engineering, Purdue University, West Lafayette, Indiana 47907*
*5 Department of Chemistry and Biochemistry, University of Notre Dame, Notre Dame, IN 46556, USA*
*6 Department of Chemical and Biomolecular Engineering, University of Notre Dame, Notre Dame, IN 46556, USA*
**jhaug@nd.edu*

Abstract: Plasmonic hyperbolic metasurfaces have emerged as an effective platform for manipulating the propagation of light. Here, confined modes on arrays of silver nanoridges that exhibit hyperbolic dispersion are used to demonstrate and model a super-resolution imaging technique based on structured illumination microscopy. A spatial resolution of ~75 nm at 458 nm is demonstrated, which is 3.1 times better than an equivalent diffraction limited image. This work emphasizes the ability to engineer the properties of confined optical modes and to leverage those characteristics for applications in imaging. The results of this work could lead to improved approaches for super-resolution imaging using designed sub-wavelength structures.

Keywords: hyperbolic metasurface, plasmonics, structured illumination microscopy, super resolution

## 1. Introduction

The desire to study and observe cellular and dynamic features of biological phenomena is a driving force behind improving the resolution limits of optical microscopy. For traditional optical microscopy, the resolution, $\Delta x$, is limited by the wavelength of light, $\lambda$, and the numerical aperture, NA, of the objective lens: $\Delta x \sim \lambda/(2\text{NA})$ [1]. For a typical imaging setup (NA = 1.4) and $\lambda$= 550 nm, a spatial resolution of approximately 220 nm can be achieved. This resolution is insufficient to observe many important sub-cellular structures, which exhibit spatial extents on the order of tens of nanometers or less [2]. To improve the spatial resolution, light with shorter wavelengths can be used; however, using this high-energy light to illuminate biological cells can damage or kill the specimens, limiting the ability to observe dynamical processes. Additionally, effects such as photobleaching can lead to deterioration of the cells, therefore masking their behavior in a natural environment [3,4]. To overcome the resolution limits in traditional imaging instruments without resorting to shorter wavelength light, a variety of super-resolution techniques have been proposed and demonstrated [5,6]. These include but are not limited to stochastic optical reconstruction microscopy (STORM) [7], hyperlens imaging [8,9], photoactivated localization microscopy [10], near-field optical scanning microscopy (NSOM) [11], stimulated emission depletion (STED) [12], and structured illumination microscopy (SIM) [13].

These techniques vary in both their conceptual approach and design implementation in achieving super-resolution images. STED techniques rely on forced stimulated emission of fluorophores in a region to create a smaller effective excitation point spread function (PSF) and then scanning that area to generate the super-resolution image. As a result, when imaging larger fields of view the imaging speed decreases rapidly, and this technique may demand specialized fluorophores like Atto dyes [14]. On the other hand, localization techniques like STORM achieve super-resolution images by slowly controlling the activation of fluorophores in a single molecule over many iterations to determine the positions of the molecules. Photoactivated localization techniques utilize specific photoreversible or photoswitchable fluorophores and generally have a tradeoff between collection and computational analysis time. Near-field scanning techniques like NSOM generally rely on leveraging fields from subwavelength apertures illuminating a sample. While NSOM is popular in small sample areas for surface or small depth of field measurements, when trying to image wider areas, the scan time becomes exceedingly long.

Compared to other super-resolution techniques, structured illumination microscopy (SIM) is of particular interest for biological applications because in addition to producing super-resolution images, SIM is a wide-field imaging technique that touts both compatibility with a wide array of fluorophores and the capability of higher imaging speeds that are needed for observation of cellular processes. SIM and SIM-variants have already been applied to *in vivo* biological imaging applications in a variety of animal cell samples, like zebrafish eyes, rabbit jejunum, and human lung fibroblasts [15,16].

The underlying principle of SIM is to capture high spatial frequency information by leveraging the Moiré effect between a structured illumination pattern and the object under investigation. The Moiré effect is a result of interference between two oscillating optical fields, which results in a low-spatial frequency pattern from the interference of the two higher spatial frequency inputs [17,18]. Standard SIM can improve the spatial resolution roughly two-fold compared to the diffraction limit [19,20]. Variants of SIM, such as plasmonic structured illumination microscopy (PSIM) and localized plasmonic structured illumination microscopy (LPSIM), rely on the same principles, but leverage different optical phenomena to generate the illumination patterns. These approaches have achieved nearly 3x improved spatial resolution compared to traditional diffraction-limited wide-field images [21-24]. Some variants of SIM, such as Blind SIM, emphasize different ways of implementing the reconstruction algorithm in order to allow for more lenient criteria on the illumination patterns used [25]. The Blind SIM algorithm utilizes an iterative approach based on the conjugate gradient method for reconstruction which allows for the use of even unknown speckle patterns as the illumination pattern. It has also been adapted and implemented experimentally for LPSIM by Ponsetto [26].

In this work, we numerically demonstrate how improved super-resolution imaging can be achieved by engineering the effective index of confined optical modes. To achieve this, we use a metasurface comprising coupled silver (Ag) nanoridge arrays that are engineered to exhibit hyperbolic dispersion. Optical modes excited on the metasurface possess a large wavevector, $k_{hyp}$. When localized, these high-$k$ modes produce standing waves where the phase of the field distribution can be controlled via the incident angle of the illuminating source, allowing these hyperbolic modes to be utilized in SIM.

## 2. Structured Illumination Microscopy

In SIM, super-resolution images are constructed from multiple diffraction-limited images. Each diffraction-limited image is the result of the Moiré fringes produced and the collecting optics, where only light with wavevectors within the passband of the optical transfer function (OTF) can be detected. The OTF is represented in reciprocal space in Figure 1 as a dashed, grey circle of radius $k_{cutoff}$, where $k_{cutoff}$ represents the maximum wavevector that can be resolved and is inversely related to the minimum spatial resolution that can be imaged in the far-field. Typical values of $k_{cutoff}$ lie close to ~$2NA/\lambda$. In SIM, wavevectors beyond $k_{cutoff}$ are accessible through interference between an object and a structured illumination pattern with wavevectors of magnitude $k_{illumination}$. Thus, the maximum wavevector contributing to the reconstructed image can be as large as $k_{cutoff}+k_{illumination}$. The super-resolution image is obtained via a reconstruction algorithm that combines multiple overlapping images with different illumination phases and orientations. The resulting resolution is proportional to $1/(k_{cutoff}+k_{illumination})$. When $k_{cutoff}\approx k_{illumination}$, the resolution improvement is close to a factor of 2. To further enhance the resolution, illumination patterns with large spatial wavevectors are needed. As mentioned previously, PSIM and LPSIM leverage the large wavevectors associated with surface plasmon polariton (SPP) and localized surface plasmon polariton (LSPP) modes to achieve their respective resolution enhancements.

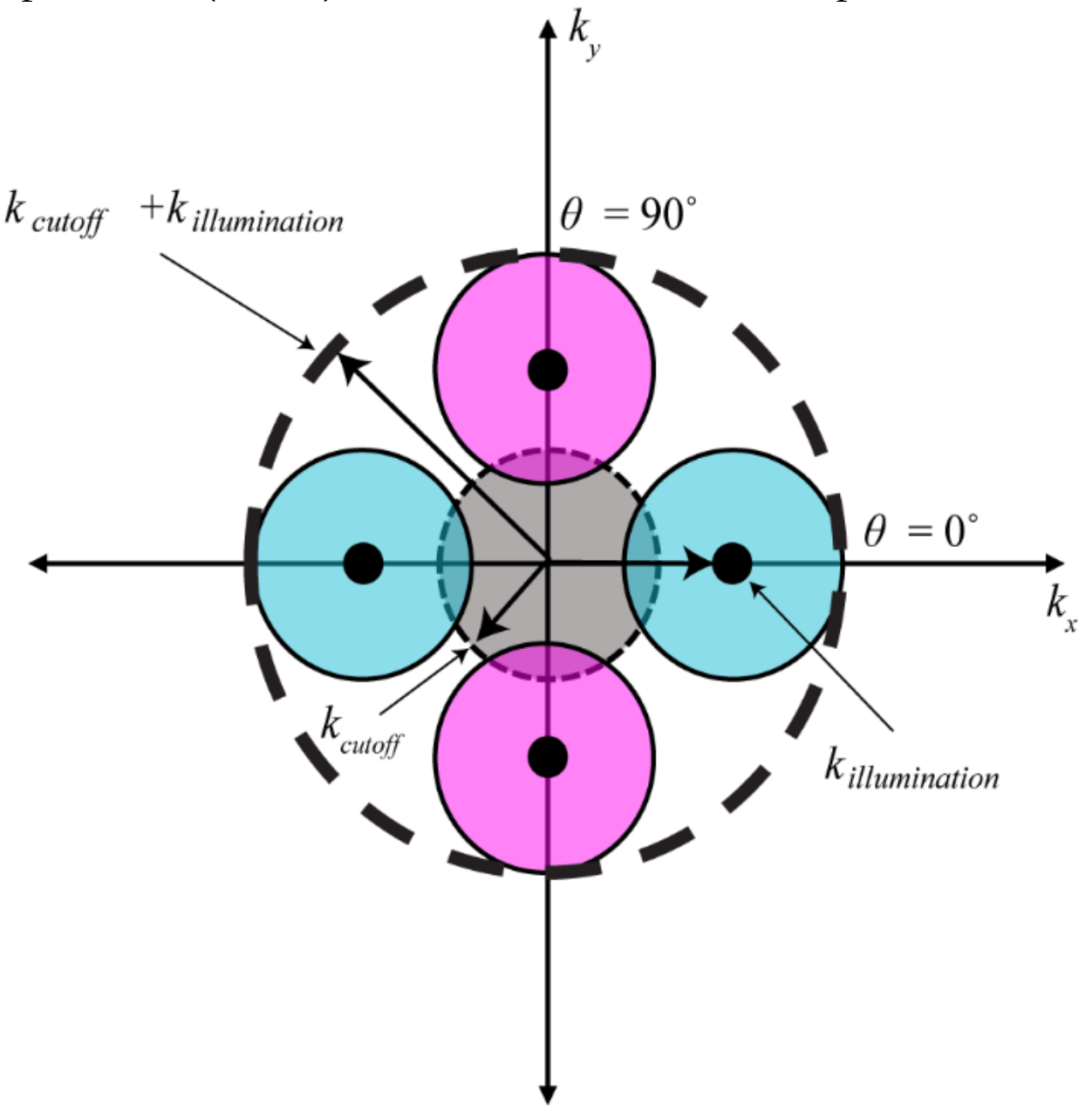

Fig. 1. The gray circle of radius $k_{cutoff}$ corresponds to the diffraction-limited PSF using conventional microscopy, and $k_{illumination}$ is the spatial frequency of the illumination pattern used. At different angles of orientation of the illumination pattern $\theta$, the direction of the PSF frequency shift changes in Fourier space correspondingly.

Plasmonic structured illumination microscopy adapts traditional SIM techniques while utilizing excited SPPs for the structured illumination pattern in the imaging process [27]. In comparison to traditional SIM techniques, the large wavevectors associated with surface plasmons ($|k_{sp}|$) have improved the image resolution by as much as 2.6x with a single illumination wavelength of 532 nm [28]. More recently, building upon PSIM, Ponsetto et. al experimentally demonstrated LPSIM utilizing LSPPs excited on metallic nanodisc arrays to surpass the previously demonstrated resolution limit of PSIM, achieving a spatial resolution of 75 nm, an improvement of ~3x with an illumination wavelength of 488 nm [29,30]. Although utilizing structured illumination patterns of different origins, SIM, PSIM, and LPSIM, all utilize the Moiré effect and reconstruction algorithms to achieve super-resolution imaging. Creating structured illumination patterns using other optical modes, could enable additional improvements in the resolution or flexibility of the approach.

## 3. Hyperbolic Metasurfaces

Metamaterials and metasurfaces are variants of artificial materials realized via engineered subwavelength structures that exhibit optical characteristics typically unseen in naturally occurring materials such as extremely large refractive indices [31], negative refraction [32], and unique electrical permittivities [33-35]. Metasurfaces have attracted attention recently due to their extensive applications in areas like wave front engineering [36], cloaking [37-39], polarization control [40,41], and imaging [42,43]. Subwavelength metallic grating metasurfaces, such as those employed here, have been shown through simulation and experiment to support optical surface modes that exhibit hyperbolic dispersion [44].

Our metasurface structure consists of Ag ridges arrays surrounded by water as shown schematically in Figure 2. Each ridge is 90 nm wide, 100 nm tall, and has length $\ell$. The ridges are separated both laterally and longitudinally by 90 nm. Water is used in the analysis, as it is a common medium for most biological cells. Our ridge structures exhibit high field intensities along their top surfaces which can facilitate the excitation of phenomena within cells and cell walls placed on top of them. The ridge width and lateral separation are engineered to achieve hyperbolic dispersion in the spectral region near $\lambda_0$ = 458 nm, a wavelength selected because it corresponds to the fluorescence excitation maximum of flavins, which are important mediators of electron transfer as part of metabolism. Most bacterial species are known to incorporate flavin adenine dinucleotide (FAD) and/or flavin mononucleotide (FMN) as part of energy generating cascades in their membranes, which can serve as a prototypical bacteria for studying metabolic path-ways.

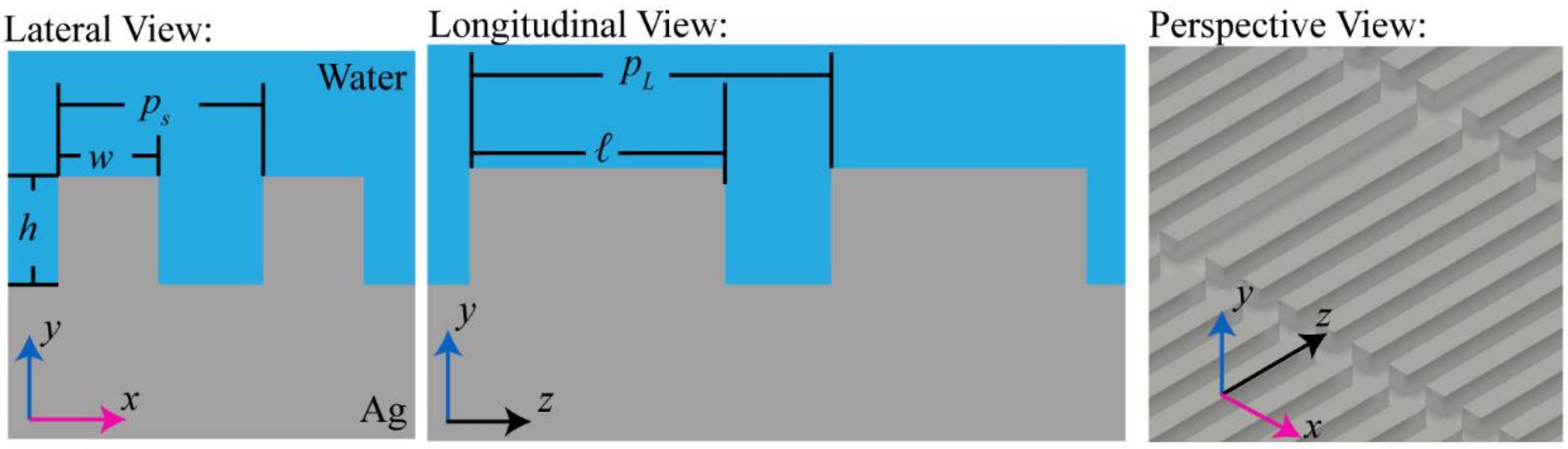

Fig. 2. The nanoridge geometry studied in both lateral (left), $x$, and longitudinal (middle), $z$, views, where $w$ is the width, $h$ is the height, and $\ell$ is the length of the ridge. The periodicity in the lateral direction is $p_s$ and in the longitudinal direction is $p_L$. The Perspective View shows an isometric view of the ridges with the corresponding coordinate system also provided (right).

Figure 3 (A) shows isofrequency curves for the hyperbolic mode for two different wavelengths ($\lambda_0$ = 458 nm and 750 nm). Here, $w$ = 90 nm, $h$ = 100 nm, and $p_s$ = 180 nm (see Fig. 2), and the top semi-infinite space is filled with water. For these calculations, the permittivity of silver is interpolated from Ref. 45 [45]. At $\lambda_0$ = 458 nm, the isofrequency curve exhibits a maximum at $k_x = \pm\pi/p_s$ and a minimum at $k_x = 0$. This behavior is characteristic of a hyperbolic metasurface where the wavevector can become large with increasing $k_x$, and wave propagation is strongly anisotropic [46]. In comparison, for $\lambda_0$ = 750 nm the dispersion exhibits a maximum at $k_x = 0$ and a minimum at $k_x = \pm\pi/p_s$, the opposite behavior compared to $\lambda_0$ = 458 nm. Figure 3(B) and (C) show the real and imaginary parts of the effective modal index of the same modes, where the effective modal index, $n_{eff}$, is given by $k_z = n_{eff} k_0$. The electric field intensity distribution obtained using COMSOL Multiphysics eigenmode analysis is shown in Figure 3(D).

The real and imaginary parts of the effective modal index for SPPs excited at a silver-water interface at $\lambda_0$ = 458 nm are also shown in Figure 3 for comparison. To calculate the SPP modes at the silver-water interface, we use the same permittivity for silver and the dispersion is given by $|k_{spp}| = |k_{photon}|\,[(\varepsilon_m\,\varepsilon_d)/(\varepsilon_m+\varepsilon_d)]$^(1/2) where $|k_{photon}|$ and $|k_{spp}|$ are the magnitude of the wavevectors for a free-space photon and the SPP, and $\varepsilon_m$ and $\varepsilon_d$ are the frequency-dependent permittivity of the metal and dielectric, respectively [48]. Figure 3(B) compares the real part of the effective model index for modes on the nanoridges surrounded by water (blue) and SPPs at a silver-water interface (black). The effective index for the coupled mode supported on the nanoridges at $k_x$=0 is approximately 30% greater than the effective index for SPPs at a silver-water interface. The difference is even larger for $k_x \neq 0$. Standing waves can be created in both systems using structures with finite lengths, which can serve as illumination patterns in SIM. The higher effective index for modes on the nanoridges will result in illumination patterns with higher spatial frequencies than those obtained using SPPs.

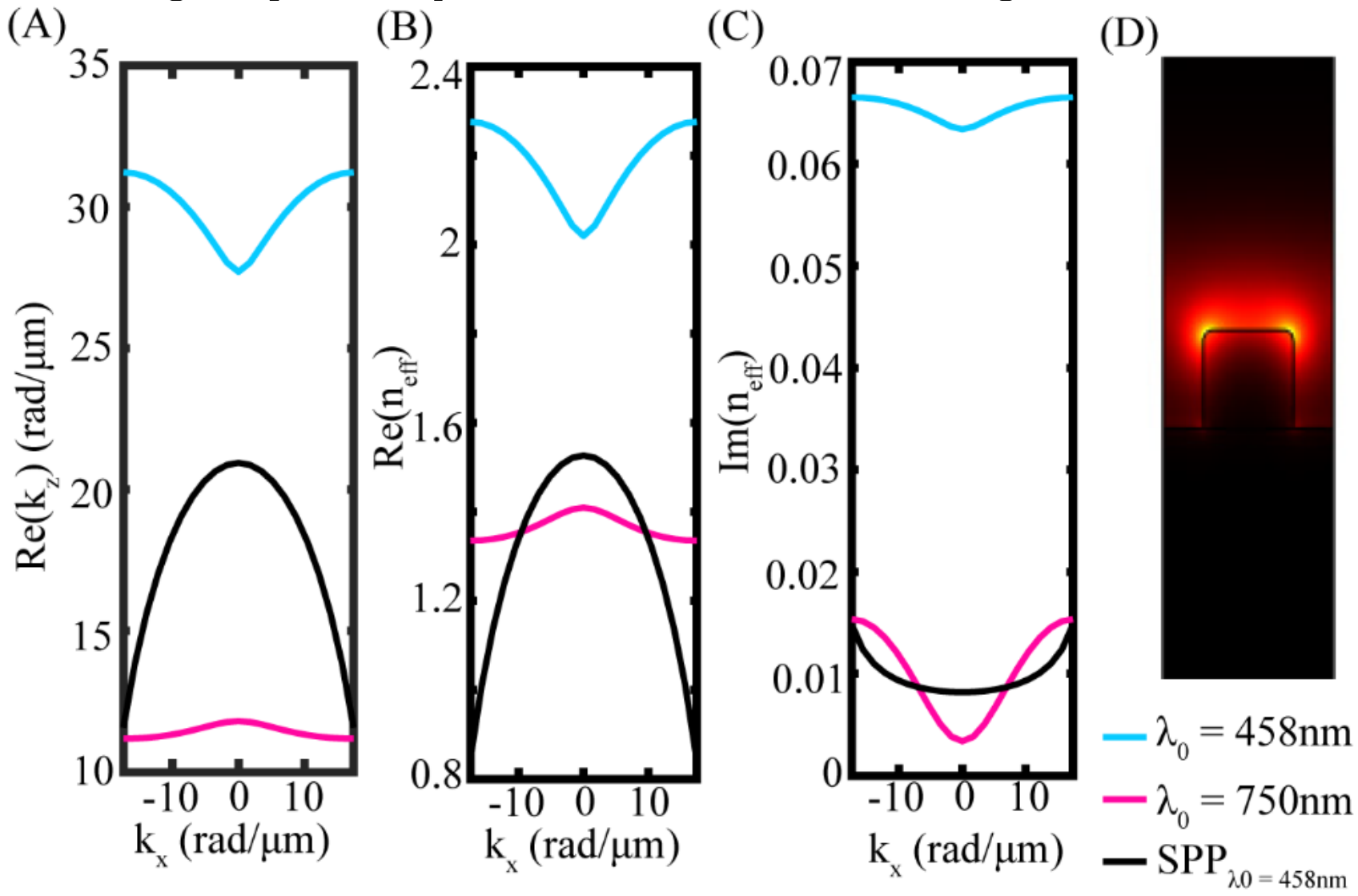

Fig. 3. (A) Real part of the wavevector in the z-direction $k_z$ for SPPs propagating along silver-water interface (black) at the top of the silver nanoridges $\lambda_0$ = 458 nm (blue) and $\lambda_0$ = 750 nm (pink) with ($w$ = 90 nm, $h$ = 100 nm, and $p_s$ = 180 nm). (B) Real and (C) imaginary parts of the effective modal index for SPPs propagating along silver-water interface (black) at the top of the silver nanoridges $\lambda_0$ = 458 nm (blue) and $\lambda_0$ = 750 nm (pink) with ($w$ = 90 nm, $h$ = 100 nm, and $p_s$ = 180 nm). The effective modal index is related to the wavevector of the propagating mode through the relation $n_{eff} = k_z/k_0$. The dispersion is hyperbolic for Re($n_{eff}$) when $\lambda_0$ = 458 nm . (D) Calculated electric field mode profile obtained using COMSOL eigenmode analysis for $\lambda_0$ = 458 nm at $k_x$ = 0.

We demonstrate the effect that the geometry of the ridge has on $n_{eff}$ in Fig. 4. Figure 4(A) shows that for our ridge structures at $\lambda_0$ = 458 nm, when setting $w$ = 90 nm and $p_s$ = 180 nm and varying the height $h$ from 60 to 120 nm there is a monotonic increase in the resulting effective modal index of the hyperbolic mode. Conversely, as shown in Figure 4(B), when varying the width of the ridge, $w$, from 60 nm to 120 nm with $h$ = 100 nm and $p_s$ = 180 nm, there is a local minimum in the magnitude of $n_{eff}$. By altering geometric parameters of the ridges one can easily engineer the dispersion of the hyperbolic mode to fit different design specifications.

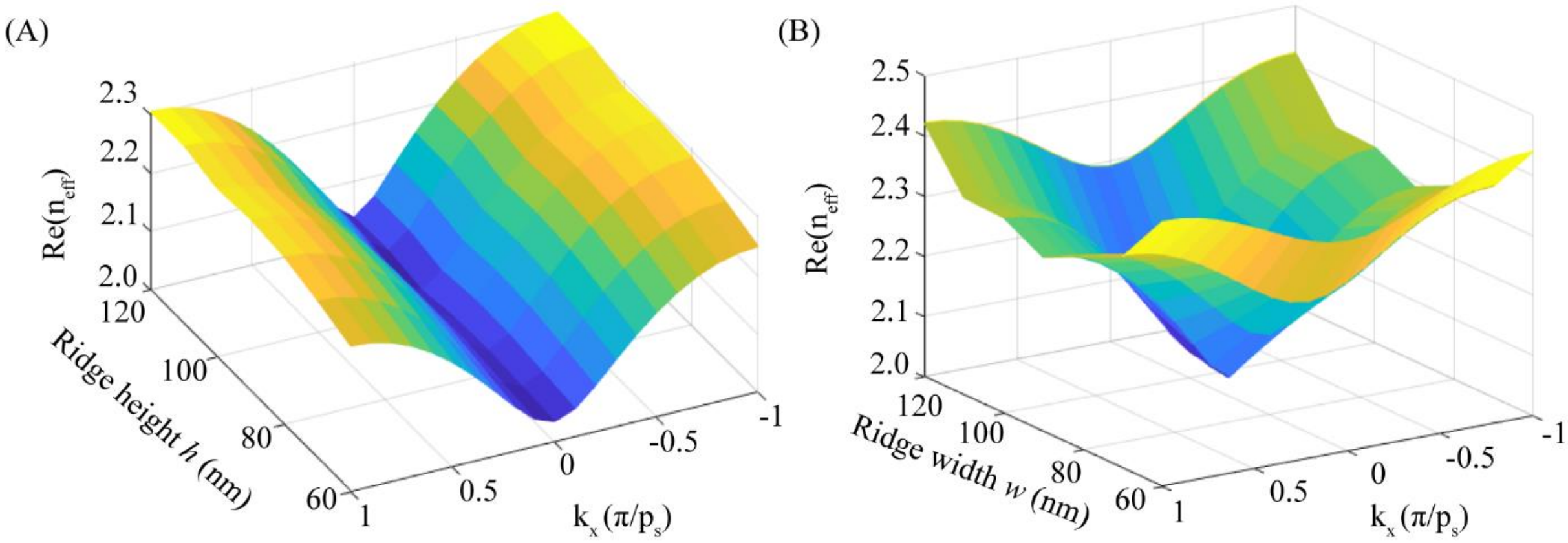

Fig. 4. (A) Real part of the effective modal index dispersion ($n_{eff}$ vs. $k_x$) for the hyperbolic mode present along silver nanoridge arrays as you alter the height of the ridges h and set $\lambda_0$ = 458 nm, $w$ = 90 nm, and $p_s$ = 180 nm. (B) Real part of the effective modal index dispersion ($n_{eff}$ vs. $k_x$) for the hyperbolic mode present along silver nanoridge arrays as you alter the width w of the ridges w and set $\lambda_0$ = 458 nm, $h$ = 100 nm, and $p_s$ = 180 nm.

## 4. Confined Hyperbolic Ridge Modes

At optical frequencies, metallic nanorod structures can act as one-dimensional Fabry-Perot resonators for confining surface plasmons [47]. By modifying the length of the nanowire and the underlying substrate, the resonant mode frequency and electric field distribution can be engineered [48,49]. Similarly, the length of the coupled ridges, ℓ, determines the frequency of supported resonant modes. We simulate arrays of silver nanoridges of varying length illuminated by plane waves using the commercial software package Lumerical.

Here, we vary only the length of the ridges in the array, keeping the other parameters constant at $h$ = 100 nm, $w$ = 90 nm, $p_s$ = 180 nm, $p_L$ = (90 nm + $\ell$), and $\lambda_0$ = 458 nm, Fig. 2. Figure 4 shows the calculated electric field magnitude along a ridge in the array for three different antenna lengths. As expected, increasing the length of the antenna results in additional antinodes. For two counter-propagating waves with wavelength $\lambda$, the resulting interference fringe peak spacing corresponds to $\lambda/2$ [50,51]. Similarly, the expected distance between antinodes in the electric field magnitude along the ridges is related to the effective modal index, $n_{eff}$, calculated using COMSOL through the expression $\Delta z = \lambda_0/(2n_{eff})$.

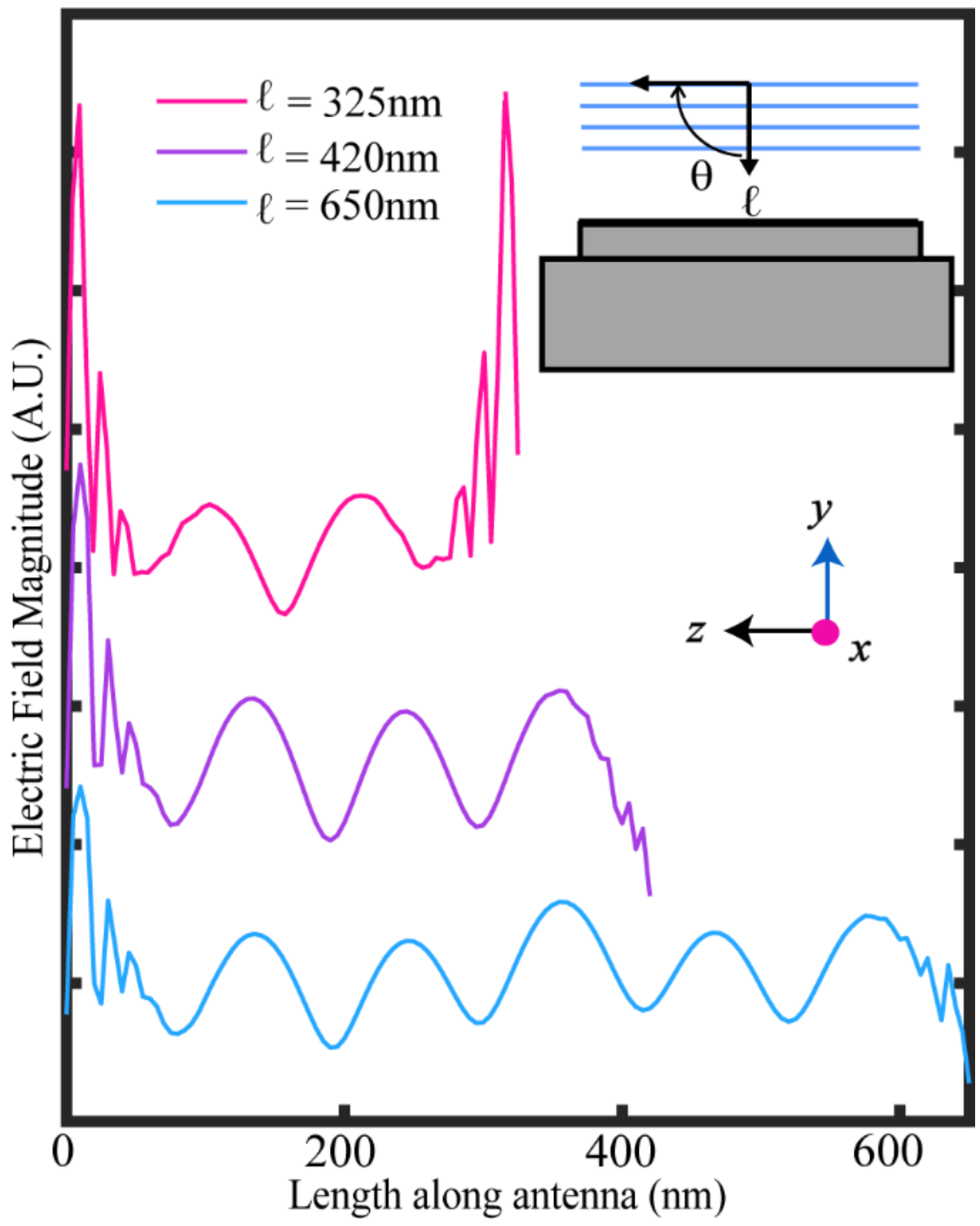

Fig. 4. Electric field magnitude plotted along the length of three different lengths of nanoridges $\ell$ (pink $\ell$ = 325nm, purple $\ell$ = 420 nm, blue $\ell$ = 650 nm) taken 5 nm above one of the top corners of the ridge, offset vertically for visibility. As the length of the nanoridges increases, the number of antinodes present along the top length of the nanoridges also increases. The parameters of the nanoridges are held constant at $w$ = 90 nm, $h$ = 100 nm, $p_s$ at 180 nm, $p_L$ = $\ell$+90 nm, and the incident light is centered at $\lambda_0$ = 458 nm with an incident angle $\theta$ = 4˚. A depiction of the illumination scheme is provided in the inset to the right of the figure, the incident light is p-polarized in this scheme.

To confirm that the mode present along the nanoridge arrays is associated with the hyperbolic mode identified in the eigenmode calculations, we calculate the spatial frequency components of the electric field magnitude using Fourier analysis, Figure 5(A). The resolution of the Fourier spectrum in our analysis is limited by the length of the ridge length. The eigenmode calculations predict an effective modal index of $n_{eff}$ = 2.019 for $k_x$ = 0. Treating the coupled nanoridge arrays of finite length as Fabry-Perot resonators with the calculated effected modal index, the predicted spacing between antinodes in the electric field magnitude is ~115 nm; this is in agreement with the field amplitude plots shown in Figure 4. In Figure 5(A) a peak at 55 rad/μm is observed, which arises from the periodicity of the electric field magnitude of the standing wave (Figure 4 and Figure 5(B)). The analytic frequency peak location would equate to $2\pi/\Delta z$, which corresponds to 54 rad/μm for $\Delta z$=115 nm. The vertical dashed line in Figure 5(A) represents this frequency in the Fourier spectrum, showing excellent agreement between the eigenmode and plane-wave excitation calculations and supporting the assignment of the excited electric field magnitude. The Fourier spectrum for confined SPP modes for 672 nm long antennas is also shown for comparison as to ensure that the resolution is consistent between the nanoridge and SPP analysis. As expected, the peak in the Fourier spectrum occurs at lower spatial frequencies. In the inset of Figure 5(A), we also show that zero padding the electric field magnitude used in our Fourier analysis yields a similar difference in peak location for our nanoridge and SPP simulations.

In order to utilize these confined hyperbolic modes in SIM, a mechanism for controlling the phase of the standing wave illumination pattern is needed. To demonstrate that the angle of incidence can be used to exert phase control over these hyperbolic modes, we calculate the electric field magnitude for several incident angles of 672 nm long antennas, ranging from 1-8°, as shown in Figure 5(B). The phase of the electric field along the antennas clearly shifts. Figure 5(C) shows the phase of the Fourier transform for each of the incident angles, demonstrating a 120° shift as the incident angle is varied. In traditional SIM ideally one is able to achieve a total shift of $4\pi/3$ or 240°, which we are able to achieve through varying the incident angle from -8° to 8°.

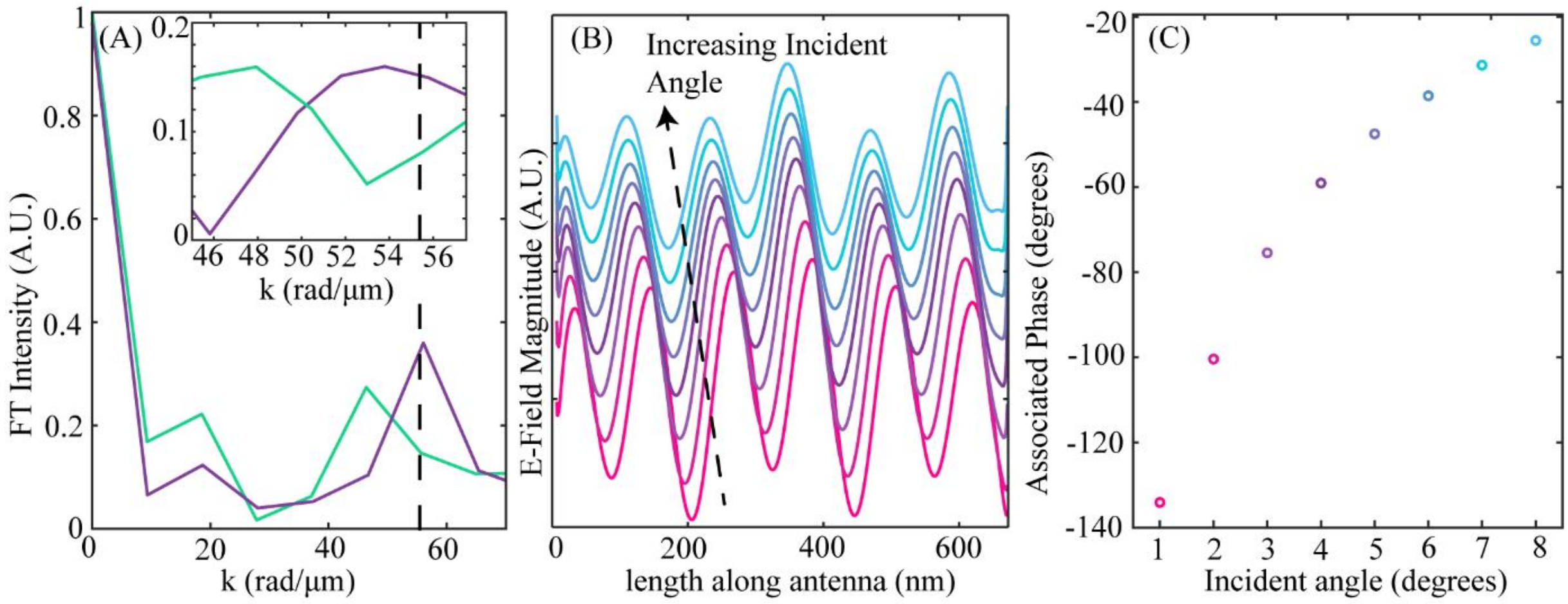

Fig. 5. (A) Fourier spectrum of the electric field plot along the length of an $\ell$ = 672 nm nanoridge (purple) at $\lambda_0$ = 458 nm at $\theta$ = 4 degrees and a water-silver SPP (green) with the gray dashed line representing the frequency corresponding to the expected frequency of the $k_x$ = 0 value for $n_{eff}$ value from the mode analysis simulations shown in Fig. 3. The inset shows the spectra from the range $k$ = 46-57 rad/µm when the corresponding electric field profiles are zero padded. (B) The electric field magnitude plotted along the length of $\ell$ = 672 nm nanoridges at $\lambda_0$ = 458 nm at 8 different incident angles (1-8 degrees, following the pink to blue color transition), staggered vertically for visibility, taken 5 nm above the top corner structure. (C) The phase of the Fourier spectrum of the electric fields from (B) at the peak located near the expected hyperbolic frequency represented by the gray dashed line in (A).

## 5. SIM Reconstruction

As a demonstration of the resolution enhancement possible using coupled nanoridge arrays, we looked to use the results of the Lumerical FDTD simulations with a slight variant of the Blind-SIM technique coined as the delta-sampling reconstruction method by Ponsetto in Ref. 26. Similar to Blind-SIM, this method utilizes an iterative algorithm based on the conjugate gradient method and removes the necessity of known sinusoidal illumination patterns compared to standard SIM techniques. Ponsetto demonstrated this method as a viable reconstruction method for LPSIM utilizing metallic nanopillar structures both in simulations and experimentally. To demonstrate the use of localized modes on metasurfaces for super-resolution imaging, we also perform Blind-SIM using the electric field pattern of our ridge arrays in the *x-z* plane taken 5 nm above the ridge arrays, shown in Fig. 7(B). In our Blind-SIM simulations, the electric field patterns corresponding to the electric field plots for incident angles 1-8° shown in Fig. 5(B) are used. Here we also use a 1024x1024 pixel image with each pixel representing 1 nm as the image parameters, NA = 1 at $\lambda_0$ = 458 nm as our optical parameters, and 10 nm radius quantum dots as the object patterns.

We demonstrate Blind-SIM reconstruction for our ridge illumination pattern (Fig. 7(A)) for three quantum dot arrangements. The first in Fig. 7(A-D) shows the ability to resolve 12 randomly scattered quantum dots. The quantum dot pattern and one of the illumination patterns used are shown in Fig. 7(A) and 7(B), respectively. It can be seen that for the diffraction-limited image in Figure 7(C) there are roughly 6 large, amorphous features whereas for the superresolved image in Figure 7(D) there are 9 finer circular features. If the dots are placed too closely together or within the electric field nulls between adjacent ridges in the *z*-direction then they will not be resolved as clearly as well-spaced dots located on the ridges. Superresolution techniques often use the comparison of full-width half-maximum's (FWHM) of image features between a diffraction-limited and super-resolved image to indicate the resolution enhancement capable with the technique. In Fig. 7(E-G) we show that for the case of a singular quantum dot we can reach enhancement of 3.1x in the FWHM comparison. This aligns with our analytic expectation of ~3x from the relation for enhancement of $(\mathrm{NA}+n_{eff})/\mathrm{NA}$. For comparison analytically, the expected resolution enhancement for PSIM for a silver-water SPP at $\lambda_0$ = 458 nm would be ~2.6x. Additionally, in Fig. 7(H-J) we demonstrate the ability to resolve two quantum dots spaced 170 nm apart center-to-center.

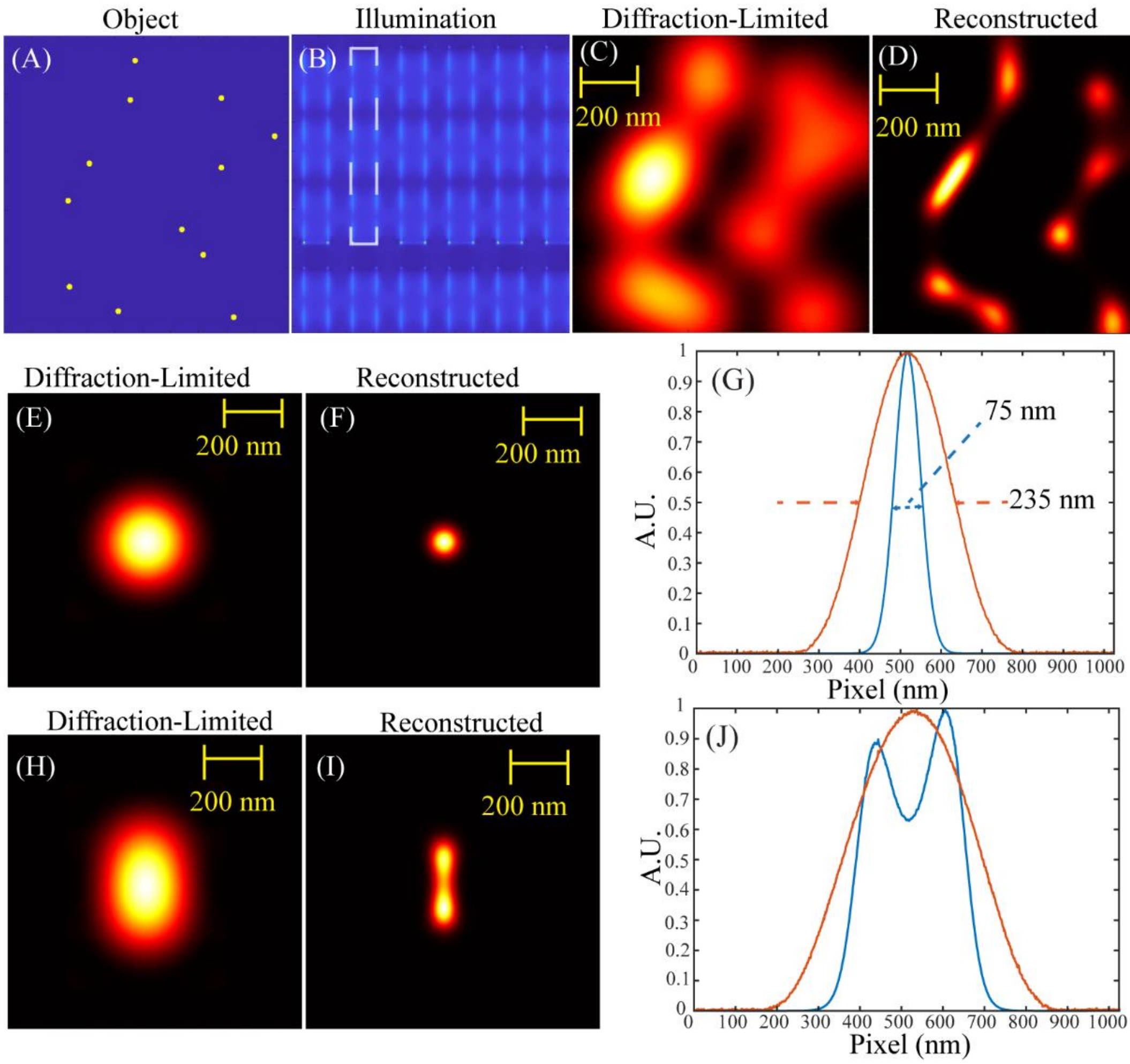

Fig. 7. Blind SIM reconstruction for a 1024x1024 image using illumination patterns corresponding to silver ridges on a silver substrate in water at $\lambda_0 = 458$ nm for incident angles 1-8°. The top row (A-D) shows an example object pattern consisting of twelve scattered 10 nm radius quantum dots, one of the illumination patterns used ($\theta = 1°$) where a ridge outline is shown with a white dashed outline, the diffraction-limited image of the quantum dots, and the reconstructed image of the quantum dots showcasing clear resolution enhancement. The middle row (E-G) shows the reconstruction and resulting FWHM for a single quantum dot as well as the intensity line profile taken horizontally through both E and F. The resulting resolution enhancement was 3.1x. The bottom row (H-J) demonstrates the ability to distinguish two quantum dots placed 170 nm apart center-to-center, with the intensity line profile taken vertically through the dot locations provided on the right that shows two clear peaks.

## 6. Conclusions

In conclusion, we have numerically demonstrated that arrays of coupled plasmonic silver nanoridges can be engineered to confine modes with hyperbolic dispersion for use in SIM. This finding is particularly important for applications requiring imaging at a specific wavelength. Our work shows how the super-resolution imaging system, in particular a structured illumination pattern, can be tailored for a particular wavelength and can improve spatial resolution in images to an equivalent degree as LPSIM while also touting engineerable dispersion based on the ridge array geometry. Our simulations show that the resolution enhancement when using these modes as structured illumination patterns is 3.1x compared to an equivalent diffraction-limited image. In addition to imaging with improved resolution, the structures explored in this work are well-poised to explore approaches in electrochemistry to exert further control over the chemical environment of the studied bacteria or biological samples.

**Funding:**

This work was supported by the U.S. Department of Energy Office of Science through grant DE-SC0019312.

**Disclosures.**

The authors declare no conflicts of interest.